\documentclass[a4paper]{jpconf}
\usepackage{graphicx}
\usepackage{graphics}

\usepackage{amsmath}
\usepackage[dvipsnames]{xcolor}

\usepackage{marvosym}
\usepackage{wasysym}
\usepackage{setspace}

\usepackage{textpos}

\begin{document}
\title{
{\bf Near and Far from Equilibrium Power-Law Statistics} 
}

\author{ \underline{{\bf T.S.~Bir\'o}}$^1$,  G.G.~Barnaf\"oldi$^1$, G.~B\'{\i}r\'o$^1$ and K.M.~Shen$^2$  }

\address{$^1$ Wigner Research Centre for Physics, Budapest, Hungary}
\address{$^2$ Central China Normal University, Wuhan, China}

\ead{biro.tamas@wigner.mta.hu}

%%%%%%%%%%%%%%%%%%%%%%%%% OWN MACROS %%%%%%%%%%%%%%%%%%%%%%%%%%%%%%%%%%
% \input{macros.tex}
%	macros.tex

\newcommand{\vs}{\vspace{3mm}}

\newcommand{\be}{\begin{equation}}
\newcommand{\ee}[1]{\label{#1} \end{equation}}
\newcommand{\ba}{\begin{eqnarray}}
\newcommand{\ea}[1]{\label{#1} \end{eqnarray}}
\newcommand{\nl}{\nonumber \\}
\newcommand{\re}[1]{(\ref{#1})}
\newcommand{\spr}[2]{\vec{#1}\cdot\vec{#2}}
\newcommand{\ave}{\overline{u}}
\newcommand{\ve}[1]{\left\vert #1  \right\vert}
\newcommand{\exv}[1]{ \left\langle {#1} \right\rangle}

\newcommand{\pd}[2]{ \frac{\partial #1}{\partial #2}}
\newcommand{\der}[2]{ \frac{{\rm d} #1}{{\rm d} #2}}   % used to be \pt
\newcommand{\pv}[2]{ \frac{\delta #1}{\delta #2}}

\newcommand{\grad}{{\vec{\nabla}}}

\newcommand{\eon}[1]{ {\rm e}^{#1}}	% used to be \ead
\newcommand{\infi}{ \int_{0}^{\infty}\limits\!}
\newcommand{\prodj}{ \prod_{j=1}^{n}\limits }
\newcommand{\sumn}{ \sum_{n=0}^{\infty}\limits\! }
\newcommand{\sumnat}{ \sum_{n=1}^{\infty}\limits\! }

\begin{abstract}
We analyze the connection between $p_T$ and multiplicity distributions in a statistical framework.
We connect the Tsallis parameters, $T$ and $q$, to physical properties like average 
energy per particle and the second scaled factorial moment, $F_2=\exv{n(n-1)}/\exv{n}^2$,
measured in multiplicity distributions. Near and far from equilibrium scenarios with
master equations for the probability of having $n$ particles, $P_n$, 
are reviewed based on hadronization transition rates, $\mu_n$, from $n$ to $n+1$ particles.
\end{abstract}

% *********************************************
%\vspace{-10mm}
%\section{Introduction}
In this paper we approach the hadronization problem in high energy physics from statistical
viewpoint.
In a balanced version of decay and growth processes a simple master equation arrives at a final
state including the Poisson, Bernoulli, negative binomial
and P\'olya distribution \cite{BiroNeda}.  Such decay and growth rates incorporate a symmetry between
the observed subsystem and the rest of a total system as a rule.
For particle physics problems $P_n$ is the probability of having
$n$ particles (or other quanta).
In an avalanche type process the simplest assumptions about elementary
rates in the master equation result in the exponential distribution with
constant rates and in the power-law tailed Waring \cite{WARING} distribution with linear
preference rates. 
In this short paper we review relevant random filling patterns in phase space and treat
thermal parameters as averages. We present master equations classified for describing dynamical
stochastic processes near and far from equilibrium, and in particular analyze stationary
distributions for large $n$. In this limit a set of coupled ordinary differential equations
is replaced by a partial differential equation. 

%%%%%%%%%%%%%%%%%%%%%%%%%%%%%%%%%%%%%%%%%%%%%%%%%%%%%%%%%%%%%%%%%%%%%%%%%%%%%%

%\section{Randomly filled Phase Space}

Following Einstein, we assume that the available phase-space volume at a given total energy, $E$,
is filled evenly in statistical equilibrium or in a stationary state. Then the blind
chance to find a part of the phase space, a.o. a single particle with energy $\omega\approx m_T-m$, 
will be given by 
$\Omega(\omega)\Omega(E-\omega)/\Omega(E)$. Beyond the one-particle phase space
factor, $\Omega(\omega)$, the rest occurs as an environmental weight factor
\be
 p(\omega) \: = \: \frac{\Omega(E-\omega)}{\Omega(E)}.
\ee{WEIGHTFACTOR}
In the simplest, idealized case the phase space volume
is just a hypersphere with radius $E$, in dimension $n$: $\Omega(E)\propto E^n$.
The above weight factor for picking out a single particle with energy $\omega$ becomes:
\be
p(\omega) \: = \: \left(1 - \frac{\omega}{E} \right)^n.
\ee{PVRATIO}
The often cited thermodynamical limit then leads to a Boltzmann--Gibbs factor
\be
p(\omega) \: \sim \: %\lim_{n\to\infty}\limits\! 
\lim_{\substack{ {E\to\infty} , {n\to\infty} \\  E/n=T}}\limits \,
\left(1-\frac{\omega}{E}\right)^n \: = \: \eon{- \, \omega/T},
\ee{TEXTBOOK}
interpreting the kinetic temperature as $T=E/n$.

However, experimentally studied physical systems are often far from the thermodynamical
limit. In {\em small systems} fluctuations can be large, and an
averaging over several millions of events is done for the histogram bins in 
obtaining $p_T$ spectra. In thermal models $\omega$ is a monotonic rising function of $p_T$.
Such an average of the above ratio, 
\be
\exv{p(\omega)} \: = \: \sumn P_n \left(1 - \frac{\omega}{E} \right)^n,
\ee{AVERp}
is interesting to be inspected for famous $n$-distributions. For the Poisson we obtain
\be
\exv{p(\omega)}^{{\rm POISSON}} = \eon{-\exv{n} \, \omega/E},
\ee{POI}
and for the negative binomial distribution (NBD)
\be
\exv{p(\omega)}^{{\rm NBD}} = \left(1+\frac{\exv{n}}{k+1}\frac{\omega}{E} \right)^{-k-1}. 
\ee{NBD}
For a general event distribution we expand eq.(\ref{AVERp})
for $\omega \ll E$ and compare the result with the Tsallis--Pareto distribution
  \be
  p(\omega) \: = \: \left(1 + (q-1) \frac{\omega}{T} \right)^{-\frac{1}{q-1}}.
  \ee{TSALLISp}
Expanding both expressions (\ref{AVERp},\ref{TSALLISp}) up to terms quadratic in $\omega$ we obtain
\be
T \: = \: \frac{E}{\exv{n}}, \qquad \mathrm{and} \qquad q = \frac{\exv{n(n-1)}}{\exv{n}^2}.
\ee{TqINTERPRET}
The parameter $T$ is the event-averaged kinetic temperature, 
while $(q-1)$ is a measure of the non-Poissonity in the multiplicity distribution. 
These findings can easily be generalized by comparing
the environmental factor (\ref{WEIGHTFACTOR}) with the Tsallis\,--\,Pareto distribution (\ref{TSALLISp}):
  \be
  \exv{p(\omega)} \: = \: \exv{\eon{S(E-\omega)-S(E)}} \: \approx \:
  \left(1 + (q-1) \frac{\omega}{T} \right)^{-\frac{1}{q-1}}.
  \ee{EINSTEINp}
Expanding  in $\omega \ll E$ both sides up to quadratic terms we obtain
\be
\frac{1}{T} \: = \: \exv{S^{\prime}(E)} = \exv{\beta}, \qquad
\qquad q = 1 - \frac{1}{C} + \frac{\Delta\beta^2}{\exv{\beta}^2}.
\ee{TqINTERPRET2}
This interprets the parameter $T$ as the inverse of the average
$\beta=S^{\prime}(E)$ and $(q-1)$ as the non-Gaussianity in $\beta$ fluctuation.
For $q=1$ the variance $\Delta\beta$  would be $1/T\sqrt{C}$ \cite{BIRO}. 
Here $\exv{S^{\prime\prime}(E)}=-1/CT^2$. 

%%%%%%%%%%%%%%%%%%%%%%%%%%%%%%%%%%%%%%%%%%%%%%%%%%%%%

%\section{Evolution via Random Changes in Particle Number}

\begin{figure}
\begin{center}
\includegraphics[height=0.25\textheight]{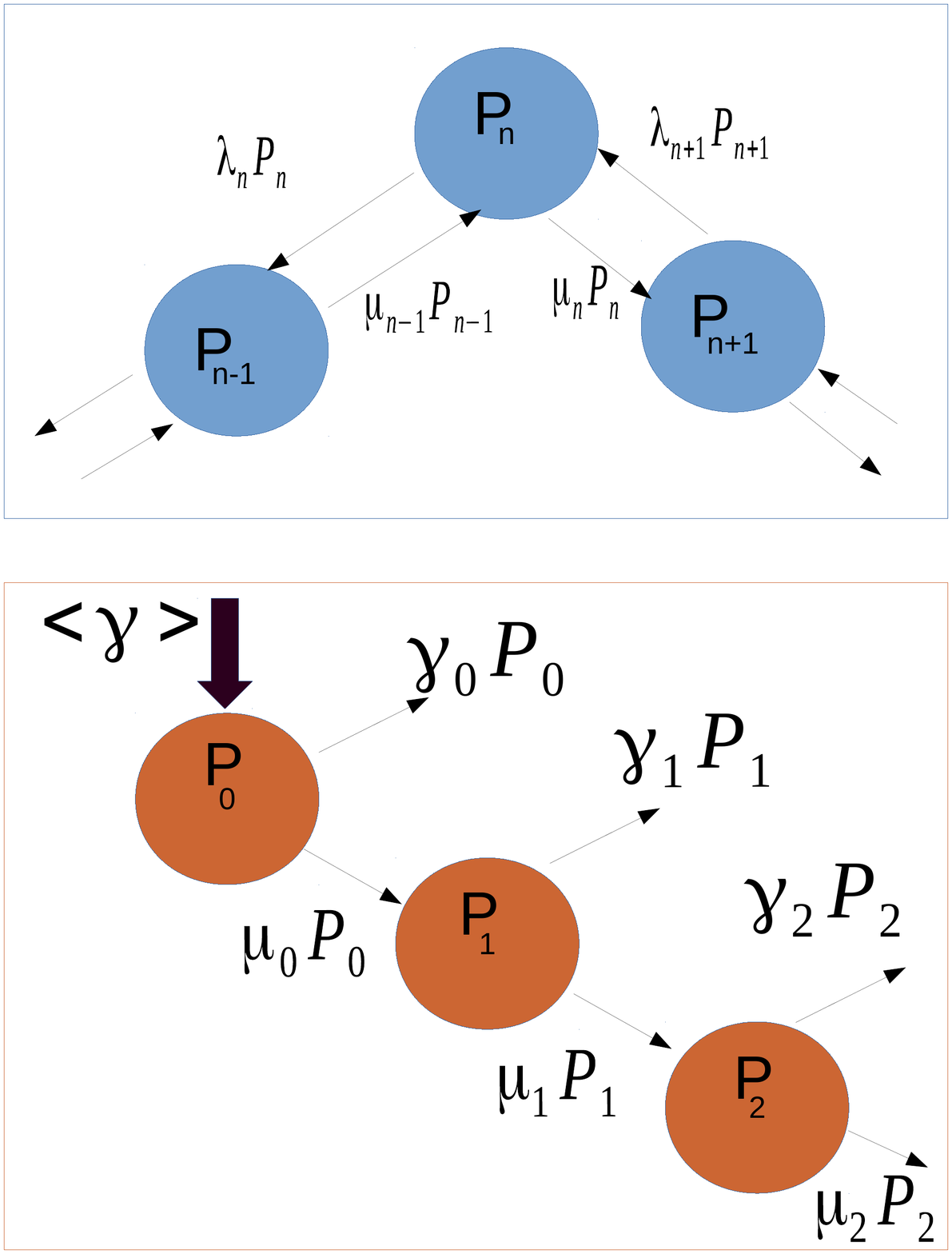}\includegraphics[width=0.18\textheight,scale=0.6,trim= -8mm -12mm 0 0]{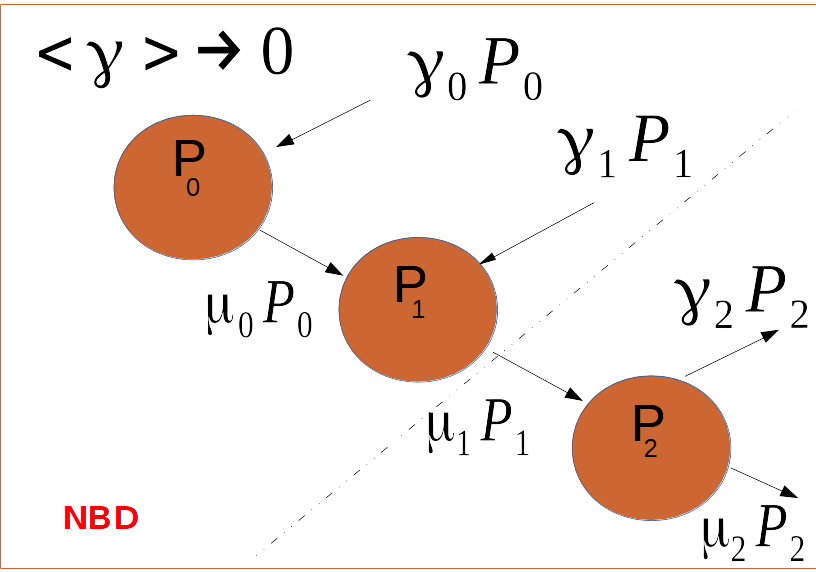}
\end{center}
\caption{\label{fig1}
Comparison of near equilibrium, diffusion like and far equilibrium, avalanche like models.
The graphs on the left side demonstrate the respective structures of the equations,
on the right side the particular one suggested to generate the NBD distribution in Ref.\cite{Japanese1997}.
}
\end{figure}

We discuss the following schemes of master equations.
The diffusion class describes near equilibrium stochastic evolution of the probability, $P_n$,
with the growth rate form $n$ to $n+1$, $\mu_n$ and the decay rate to $n-1$, $\lambda_n$:
\be
\dot{P}_n \: =  \: \left[(\lambda P)_{n+1} \, - \, (\lambda P)_n \right] 
 \, - \, \left[(\mu P)_{n} \, - \, (\mu P)_{n-1} \right]. 
\ee{MASTER}
The avalanche class describes a contesting race eventually achieving stationary branching
ratios:
\ba
\dot{P}_0 \: &=& \: \exv{\gamma} - (\gamma_0+\mu_0)P_0,
\nl
\dot{P_n} \: &=& \: \mu_{n-1} P_{n-1} \, - \, \left(\mu_n + \gamma_n \right) \, P_n \qquad {\rm for} \: n\ge 1.
\ea{EXPSCALEDMASTER}
Here the decay rate, $\gamma_n$, describes exit from the chain, a reduction in $n$ is not assumed.
Fig.\ref{fig1} depicts the difference between these classes.
%%%%%%%%%%%%%%%%%%%%%%%%%%%%%%%%%%%%%%%%%%%%%%%%%%%%%%%%%%%%%%%%%%%%%%%%%
%%%%%%%%%%%%%%%%%%%%%%%%%%%%%%%%%%%%%%%%%%%%%%%%%%%%%%%%%%%%%%%%%%%%%%%%%
By studying the stationary branching in the avalanche type dynamics, we are especially interested in
the mean aging model, where for $\forall n: \, \gamma_n=\gamma$ and therefore $\exv{\gamma} = \gamma$.
In the stationary limit: $P_n(t) \to Q_n$, and from $\dot{Q}_n=0$ one obtains 
\be
Q_n \: = \: \frac{\mu_{n-1}}{\mu_n+\gamma} \, Q_{n-1} \: = \: \cdots \: = \:
%	\frac{\gamma}{\gamma+\mu_0} \, \prod_{j=1}^{n}\limits \frac{\mu_{j-1}}{\mu_j+\gamma}.
 \frac{\gamma}{\mu_n} \: \eon{- \sum_{j=0}^n\limits\! \ln(1+\gamma/\mu_j)}.
\ee{EXPSCALEDMASTERSTAC}
Now, constant growth rates, $\mu_j=\sigma$ lead to the exponential distribution
\be
	Q_n \: = \: \frac{1}{1+\sigma/\gamma} \, \eon{-n \cdot \ln(1+\gamma/\sigma)}.
\ee{EXPSCALEDDISTR}
Linear preference growth rates \, $\mu_j = \sigma (j+b)$ \,  ($b>0$) 
lead to the Waring distribution \cite{WARING,NETWORKWARING,SCHUBERT},
\be
	Q_n \: = \: \frac{\gamma}{\gamma+b\sigma} \, \,
	\frac{\Gamma(n+b)\, \Gamma(b+1+\gamma/\sigma)}{\Gamma(b) \, \Gamma(n+b+1+\gamma/\sigma)}.
\ee{POWSCALEDDISTR}
This distribution has a power-law tail for large $n$ as $\sim n^{-1-\gamma/\sigma}$.
For negligible decay compared to the growth, $\gamma \ll \sigma$, one arrives at the Zipf distribution.
%%%%%%%%%%%%%%%%%%%%%%%%%% NBD from avalanche %%%%%%%%%%%%%%%%
It is interesting to note that it has been suggested by Osada et.al. \cite{Japanese1997} 
that the special dynamics
with $\gamma_n=\sigma(n-kf)$ and $\mu_n=\sigma f (n+k)$ leads exactly to the NBD as stationary distribution,
\be
 Q_n \: = \: \binom{n+k-1}{n} \, f^n \, (1+f)^{-n-k}.
\ee{NBDFROMAVALANCHE}
We note that for large $n$ the avalanche dynamics effectively uses the variable $x=n\Delta x$ 
and solves 
$\pd{}{t} P(x,t) \: = \:  - \pd{}{x} \left(\mu(x) \, P(x,t) \right) - \gamma(x) P(x,t)$.
The stationary distribution is
\be
{Q(x) \: = \:  \frac{K}{\mu(x)} \, \eon{-\int_0^x\limits \frac{\gamma(u)}{\mu(u)} \, du}.}
\ee{CONTISTATIONARY}
In this framework the constant rate $\mu(x)=\sigma$ leads to the exponential, and
the linear rate, $\mu(x)=\sigma (x+b)$ to the Tsallis--Pareto stationary distribution \cite{PARETO}
\be
Q(x) \: = \: \frac{\gamma}{\sigma \, b} \, \left(1+\frac{x}{b} \right)^{-1-\gamma/\sigma}.
\ee{CONTIQLINEARMU}

%\section{Experimental data}

\begin{figure}
%\center{\includegraphics[width=0.30\textwidth,scale=1.25]{TvsqBiroG.png}}
\center{\includegraphics[width=0.75\textwidth,scale=1.25]{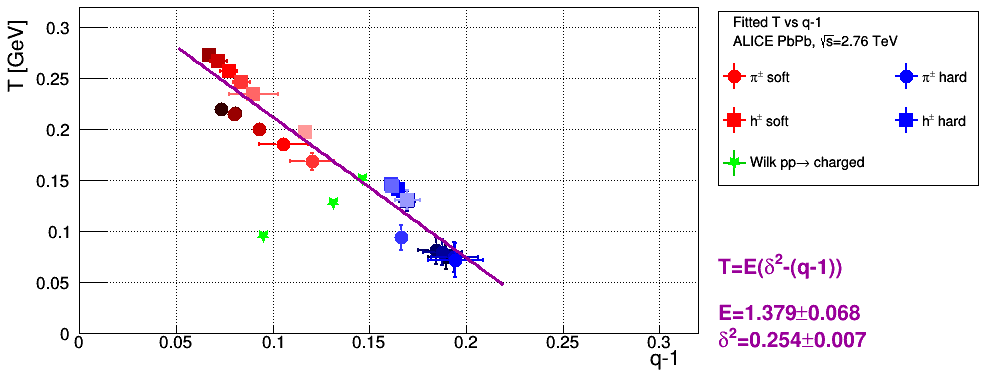}}
\caption{\label{Fig3}
 The Tsallis parameters $T$ vs $q-1$ from fits to the $2.76$ TeV ALICE PbPb data.
}
\end{figure}

A connection between $p_T$ and $n$ distributions is hinted at in finding different
$q$ values for different participant numbers in experiments \cite{ALICE,PHENIX,PHENIXsurvey2005}. 
Our model form eqs.(\ref{TqINTERPRET},\ref{TqINTERPRET2}) 
predicts $T=E(\delta^2-(q-1))$ with $E = CT \approx 1.4$ GeV and 
$\delta = \Delta\beta/\exv{\beta} \approx 0.5$. 
Fig.\ref{Fig3} shows our fit results to the LHC ALICE PbPb data at $2.76$ TeV.
Darker points belong to more central collisions. 
We also distinguish between soft ($p_T < 5$ GeV, red data points) and
hard ($p_T > 3$ GeV, blue data points) spectral parts.
For fit parameters and further details see legend.
The green stars are data from pp collision \cite{WILK}, lying on a $T=E(q-1)$ line with $E\approx 1$ GeV.
The AA points seem to favor $\delta^2=0.25$ independent of $\exv{n}$, while the pp points
$\delta^2\propto 1/\exv{n}$, meaning a constant $f=\exv{n}/k$ value in the NBD distribution.

%*******************************************************
\vspace{2mm}
{\bf Acknowledgments}:
Discussions with Zolt\'an N\'eda and Andr\'as Telcs are gratefully acknowledged.
This work has been supported by NKFIH (OTKA K 104260, 120660 projects) and
by a bilateral Chinese--Hungarian governmental project, TeT 12CN-1-2012-0016.
GGB thanks the J\'anos Bolyai scholarship of the Hungarian Academy of Science.

%%%%%%%%%%%%%%%%% BIBLIO NOTES %%%%%%%%%%%%%%%%%%%%

\end{document}